\documentclass[journal=nalefd, manuscript=letter]{achemso}

\usepackage[T1]{fontenc}
\usepackage[utf8]{inputenc}
\usepackage{graphicx}
\usepackage{mhchem} 

\title{Temperature-Dependent Spin-Orbit Torque Generation in Perpendicularly Magnetized Topological Insulator-Magnetic Multilayer Heterostructures}
\author{Soumyarup~Hait}
\email{S.Hait@leeds.ac.uk}
\author{Benjamin~A.~Brereton}
\author{Ahmet~Yagmur}
\author{Satoshi~Sasaki}
\author{Gavin~Burnell}
\author{Christopher~H.~Marrows}
\email{C.H.Marrows@leeds.ac.uk}

\affiliation[University of Leeds]
{School of Physics and Astronomy, University of Leeds, Leeds LS2~9JT, United Kingdom}

\begin{document}
\maketitle

\begin{abstract}
We report a comprehensive temperature‑dependent investigation of spin–orbit torque (SOT) generation in heterostructures comprising a perpendicularly magnetized metallic multilayer grown on top of a topological insulator (TI) epilayer. Temperature‑dependent second‑harmonic Hall measurements reveal distinct trends in the magnitude of the spin-orbit torque across the studied heterostructures. Samples incorporating Bi$_2$Se$_3$ exhibit torques reaching values of approximately $\sim 12$~mT/$10^{12}$Am$^{-2}$ at 15 K, around 5 times larger than those in a multilayer without the topological layer. The structure with a thin 2-nm Ta buffer for the multilayer shows the strongest enhancement and a pronounced increase at low temperatures, highlighting efficient spin‑current generation from the topological surface states. In contrast, the sample with a 10 nm-thick Ta spacer exhibits reduced torque efficiency, consistent with partial attenuation of spin transmission through the buffer. Systems lacking Bi$_2$Se$_3$ but containing two heavy metals (Ta and Pt) yield significantly smaller torques, around $\sim 2.5$~mT/$10^{12}$Am$^{-2}$, despite the presence of conventional spin Hall sources. These observations underscore the dominant role of TI‑derived spin–momentum‑locked currents in driving large damping‑like torques and their sensitivity to interfacial structure and buffer‑layer thickness. 

\end{abstract}

\clearpage

\section*{Introduction}

Spin–orbit torques (SOTs) have emerged as a central mechanism for electrically manipulating magnetization in next‑generation spintronic devices, offering a pathway toward ultrafast, low‑power, and scalable memory and logic technologies \cite{liu2012spin,miron2011perpendicular,shao2021roadmap,nguyen2024recent}. Their importance is particularly pronounced in magnetic multilayer (MML) heterostructures exhibiting perpendicular magnetic anisotropy (PMA), where more efficient current‑induced magnetization modification can be achieved \cite{wang2023room,ramaswamy2018recent,chen2022electrical}. PMA systems also provide enhanced thermal stability\cite{ikeda2010perpendicular} and compact device footprints\cite{peng2017interfacial}, making them attractive for applications in memory\cite{dieny2017perpendicular,ikeda2010perpendicular} and logic\cite{li2021all,debashis2018experimental}. Recent advances highlight that optimizing SOT efficiency requires careful engineering of material interfaces, symmetry breaking, and spin‑current sources, especially in multilayer structures where interfacial scattering and spin–momentum locking can dramatically modify torque generation \cite{shao2021roadmap,ramaswamy2018recent}. 

Traditionally, heavy metals (HMs) such as Pt, Ta, and W have served as primary SOT sources due to their strong spin Hall effect. However, the discovery of topological insulators (TIs) has reshaped this landscape. TIs host spin‑momentum–locked topological surface states (TSSs) that can generate exceptionally large charge‑to‑spin conversion efficiencies, often surpassing those of conventional HMs \cite{mellnik2014spin,wang2022spin,wu2019room,wang2023room,wang2015topological,choi2024highly}. Recent studies \cite{choi2024highly,dc2018room} have demonstrated that carefully engineered TI/FM heterostructures can achieve high spin–orbit torque efficiencies and low critical switching currents, even at room temperature, by maximizing current transport through the topological surface states (TSS) while suppressing bulk-state current leakage. These findings highlight the potential of TI/FM heterostructures as promising platforms for efficient SOT-driven magnetization control. 

Beyond switching, PMA magnetic heterostructures with a strong interfacial Dzyaloshinskii–Moriya interaction (DMI) can intrinsically host chiral spin textures such as N\'{e}el‑type domain walls and skyrmions \cite{ajejas2022interfacial,wang2021manipulating,cheng2023room}. When a TI is grown adjacent to such a DMI-stabilized PMA stack, the giant SOT generated by its spin–momentum-locked surface states may enable more efficient skyrmion motion and domain-wall propagation, offering a promising route toward lower-power spintronic devices. Although the present work focuses on SOT generation, the multilayer structures investigated here have previously been shown to host chiral magnetic structures in our earlier work  \cite{brereton2026tailoring}. The present study provides an important step toward understanding how SOT can be used to electrically manipulate these chiral textures. In ongoing studies, we are further investigating the formation and SOT-driven motion of skyrmions in these multilayers, with the aim of establishing a direct connection between the SOT characteristics and the dynamics of chiral magnetic textures. Such SOT-driven control of chiral textures is of particular interest for emerging technologies, including 
 racetrack‑memory \cite{fert2013skyrmions,zhang2015skyrmion}, neuromorphic computing \cite{song2020skyrmion,grollier2020neuromorphic}, or logic in-memory computing \cite{luo2018reconfigurable,zhang2020skyrmion}. Their reliable manipulation requires not only strong damping‑like torques but also precise control over torque polarity and symmetry properties that are highly sensitive to interfacial quality, buffer layer engineering, and the relative contributions of TI and HM-derived spin currents \cite{shao2021roadmap,ramaswamy2018recent}.

Studying SOTs in such exotic chiral magnetic heterostructures is inherently challenging. The physical realization of DMI and spin–orbit coupling (SOC) demands complex stacks of magnetic and non‑magnetic ultrathin films, often involving multiple angstrom‑scale repetitions to achieve the desired interfacial interactions \cite{zhao2025interlayer, alshammari2021scaling, barker2024phase}. Moreover, when TIs are incorporated, their TSSs evolve strongly with temperature, leading to unconventional transport and torque behavior \cite{noyan2025highly,wang2015topological}. Hence, a comprehensive temperature‑dependent study is essential to understand the underlying physics governing torque generation. Despite these complexities, such investigations are both scientifically and technologically significant, as they provide crucial insights into efficient spin‑current transmission and pave the way for next‑generation spintronic devices combining topological and chiral functionalities \cite{han2021topological,fert2017magnetic}.

In this work, we present a comprehensive temperature‑dependent study of SOT generation in a series of perpendicularly magnetized heterostructures grown on an epilayer of the well-known TI, Bi$_2$Se$_3$, using second‑harmonic Hall measurements across a broad temperature range. We quantitatively compare torque magnitudes originating from TIs and with those from conventional heavy metals, while systematically examining the influence of buffer‑layer thickness. Our results reveal clear distinctions in torque polarity, efficiency, and temperature scaling between TI‑based and HM‑based samples, demonstrating the unique role of TSS‑driven spin currents in chiral PMA systems. These insights advance the understanding of SOT mechanisms in complex heterostructures and highlight the promise of TI‑based platforms for future spintronic memory, neuromorphic computing, spin‑Hall oscillators, and quantum information technologies.

\section*{Results}

\subsection*{Sample details, structural and magnetic characterization}

\begin{figure}[t]
    \centering
	\includegraphics[width=6in]{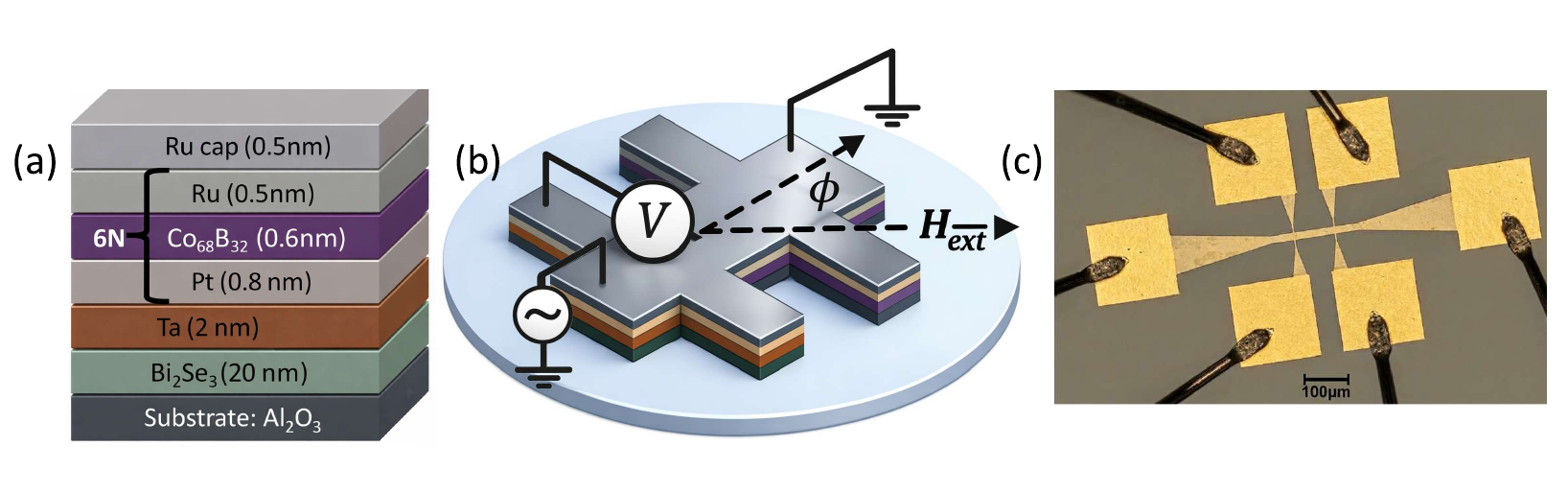}
	\caption{Experimental schematics. (a) Schematic illustration of an example TI/MMl heterostructure stack: Al$_2$O$_3$/Bi$_2$Se$_3$(20\,nm)/Ta(2\,nm)/MML. (b) Schematic representation of the Hall measurement geometry. The detailed measurement procedure is provided in the Supplementary Information. (c) Optical micrograph of the Hall bar corresponding to the sample Al$_2$O$_3$/Bi$_2$Se$_3$(20\,nm)/Ta(2\,nm)/MML. \label{fig:schematic}}
\end{figure}

The following samples were grown, characterized, and measured in this study. Further details regarding their growth and characterization can be found in our previous report \cite{brereton2026tailoring}.
\begin{enumerate}
    \item SiO$_2$/MML
    \item SiO$_2$/Ta(2\,nm)/MML
    \item SiO$_2$/Ta(5\,nm)/MML/Pt(4\,nm)
    \item Al$_2$O$_3$/Bi$_2$Se$_3$(20\,nm)/Ta(2\,nm)/MML
    \item Al$_2$O$_3$/Bi$_2$Se$_3$(20\,nm)/Ta(10\,nm)/MML
\end{enumerate}
where SiO$_2$ refers to the thermal oxide surface of a Si wafer, Al$_2$O$_3$ refers to c-plane sapphire, and the MML stack is $\left[\mathrm{Pt}(0.8\,\mathrm{nm})/ \mathrm{Co}_{68}\mathrm{B}_{32}(0.6\,\mathrm{nm})/
\mathrm{Ru}(0.5\,\mathrm{nm})\right]_6$. We have used Ta as a buffer layer for MML growth on both SiO$_2$ and Al$_2$O$_3$/Bi$_2$Se$_3$(20\,nm) substrates. The heterostructure stack is illustrated in Fig.~\ref{fig:schematic}. 

All the fabricated samples with Ta buffer exhibit smooth, well-defined interfaces and perpendicular magnetic anisotropy (PMA), achieved after extensive optimization of the growth conditions as mentioned in the supplementary information. These structural and magnetic characteristics have been thoroughly discussed in our previous report \cite{brereton2026tailoring}, supported by XRD, XRR, HAADF-STEM, AFM, and SQUID-VSM measurements. In the present work, we also performed in-plane SQUID-VSM measurements to extract the anisotropy field of the heterostructures as a function of temperature as discussed in the supplementary information. Samples were patterned into Hall bars for second harmonic measurements, illustrated in Fig.~\ref{fig:schematic}(b,c).

\subsection*{Temperature-dependent resistance and current shunting}

\begin{figure}
    \centering
	\includegraphics[width=5in]{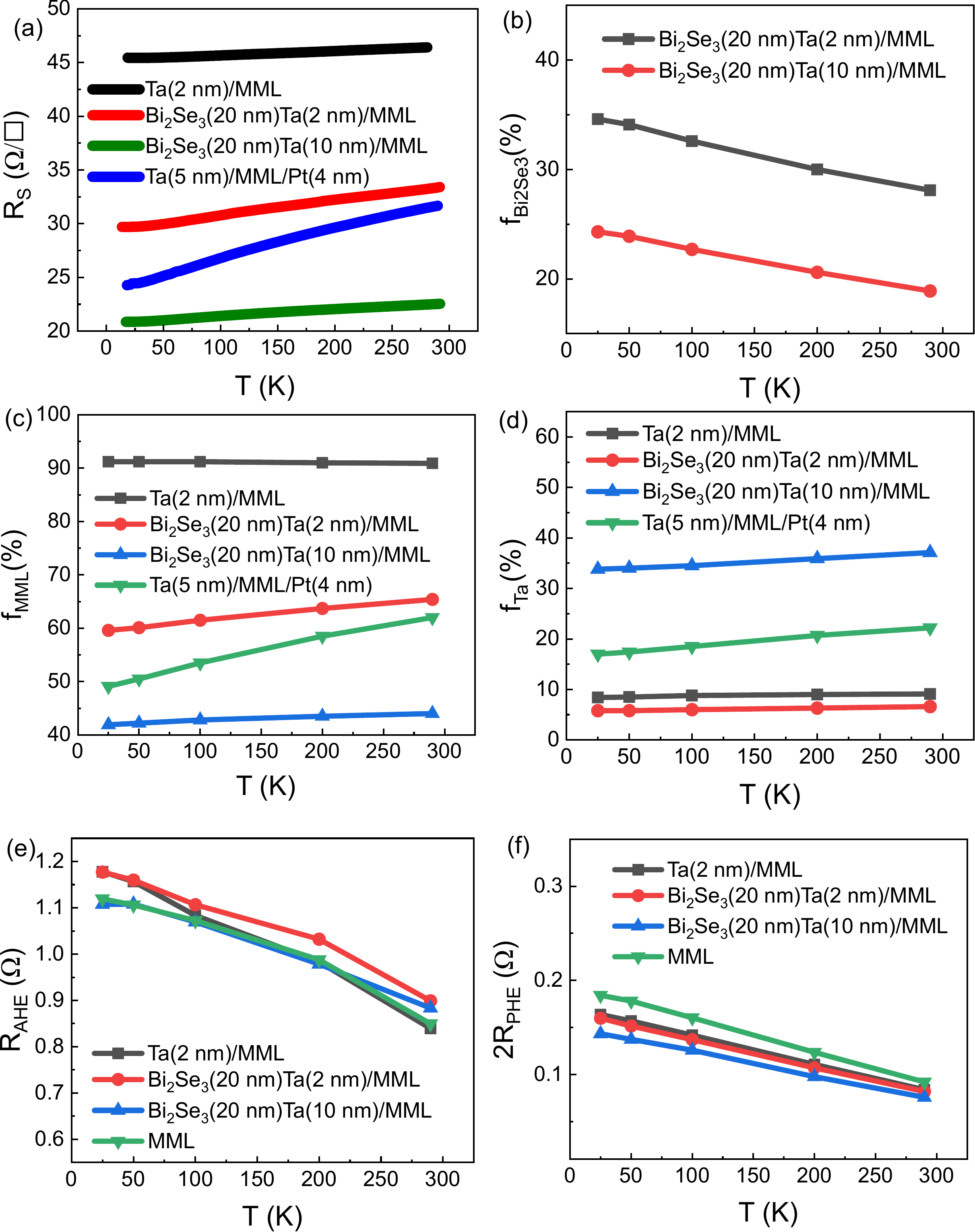}
	\caption{Temperature dependent transport properties.  (a) Temperature-dependent sheet resistance of the samples SiO$_2$/Ta(2\,nm)/MML, Al$_2$O$_3$/Bi$_2$Se$_3$(20\,nm)/Ta(2\,nm)/MML, and Al$_2$O$_3$/Bi$_2$Se$_3$(20\,nm)/Ta(10\,nm)/MML. (b--d) Fraction of the total current flowing through the  Bi$_2$Se$_3$ ($f_{\mathrm{Bi_2Se_3}}$),MML ($f_{\mathrm{MML}}$), and Ta buffer ($f_{\mathrm{Ta}}$), respectively,  as obtained from the parallel-resistor model, described in detail in the Supplementary Information.  (d--f) Temperature dependence of the  anomalous Hall resistance, and planar Hall  resistance for the heterostructures. \label{fig:shunting}}
\end{figure}

In a complex multilayer system such as the ones investigated here, it is essential to understand the temperature-dependent current shunting among the constituent layers to avoid miscalculation of key physical parameters. To this end, we measured the temperature-dependent resistance of all samples, which exhibit metallic behaviour with distinct sheet resistances $R_\mathrm{S}$ depending on the specific stack configuration, as shown in Fig.~\ref{fig:shunting}(a).

Using a parallel-resistor model, we calculated the fraction of the applied current carried by each layer, as illustrated in Fig.~~\ref{fig:shunting}(b--d). For simplicity, the magnetic multilayer (MML) is treated as a single effective conducting layer. The methodology used for the calculation is detailed in the Supplementary Information.  In the samples containing \ce{Bi2Se3}, approximately 20--35\% of the total current flows through the \ce{Bi2Se3} layer, and this fraction $f_\mathrm{Bi2Se3}$ increases as the temperature decreases. It is important to note that the quoted current density corresponds to the effective device current density rather than the current density within the TI layer itself. For the sample with the thinner Ta buffer, where Ta carries a small fraction $f_\mathrm{Ta}$ of the current (6--7\%), the \ce{Bi2Se3} layer conducts 28--35\% of the current, while the MML carries the remaining fraction $f_\mathrm{MML}$ of 60--65\%. In contrast, for the sample with the thicker Ta buffer (10~nm), the Ta layer carries a significant portion of the current (33--37\%), the \ce{Bi2Se3} layer carries 24--28\%, and the MML carries the remaining 42--44\% across the entire temperature range. As the temperature decreases, the current shunted through the \ce{Bi2Se3} increases, while the fractions flowing through the MML and Ta layers decrease correspondingly. For the sample without \ce{Bi2Se3} and with a thin Ta buffer (2~nm), nearly 92\% of the current is shunted through the MML while the rest travel through the Ta layer. In the sample containing both Ta and Pt on the bottom and top of MML, the MML carries almost (50--60\%) of the applied current, while the rest of the current flows through the two heavy metals. The current shunting fractions for the different layers are summarized in Table 2 of the Supporting Information.

\subsection*{First harmonic Hall analysis} 

Before performing the second-harmonic Hall measurements, we first carried out temperature-dependent first-harmonic Hall measurements to determine the anomalous Hall resistance ($R_{\mathrm{AHE}}$) and planar Hall resistance ($R_{\mathrm{PHE}}$). These values were corrected using the current-shunting factors obtained from the calculations described earlier. The exact shunting values are tabulated in the Supplementary Information. This step is essential, as the corrected Hall resistances directly influence the extracted torque efficiencies, as discussed in the following section.

Fig.~\ref{fig:shunting}(e) and (f) present the temperature evolution of $R_{\mathrm{AHE}}$ and $R_{\mathrm{PHE}}$, respectively, for all heterostructures studied. As expected, both Hall resistances increase as the temperature is lowered, consistent with typical magnetic behavior. Furthermore, the corrected Hall resistivities exhibit close agreement across samples, confirming the reproducibility and uniformity of the MML quality within the heterostructures. The representative AHE and PHE data for all the samples at different temperatures are shown in the Supplementary Information. 

\subsection*{Second harmonic Hall analysis for spin-orbit torque estimation} 

\begin{figure}[t]
    \centering
	\includegraphics[width=5.5in]{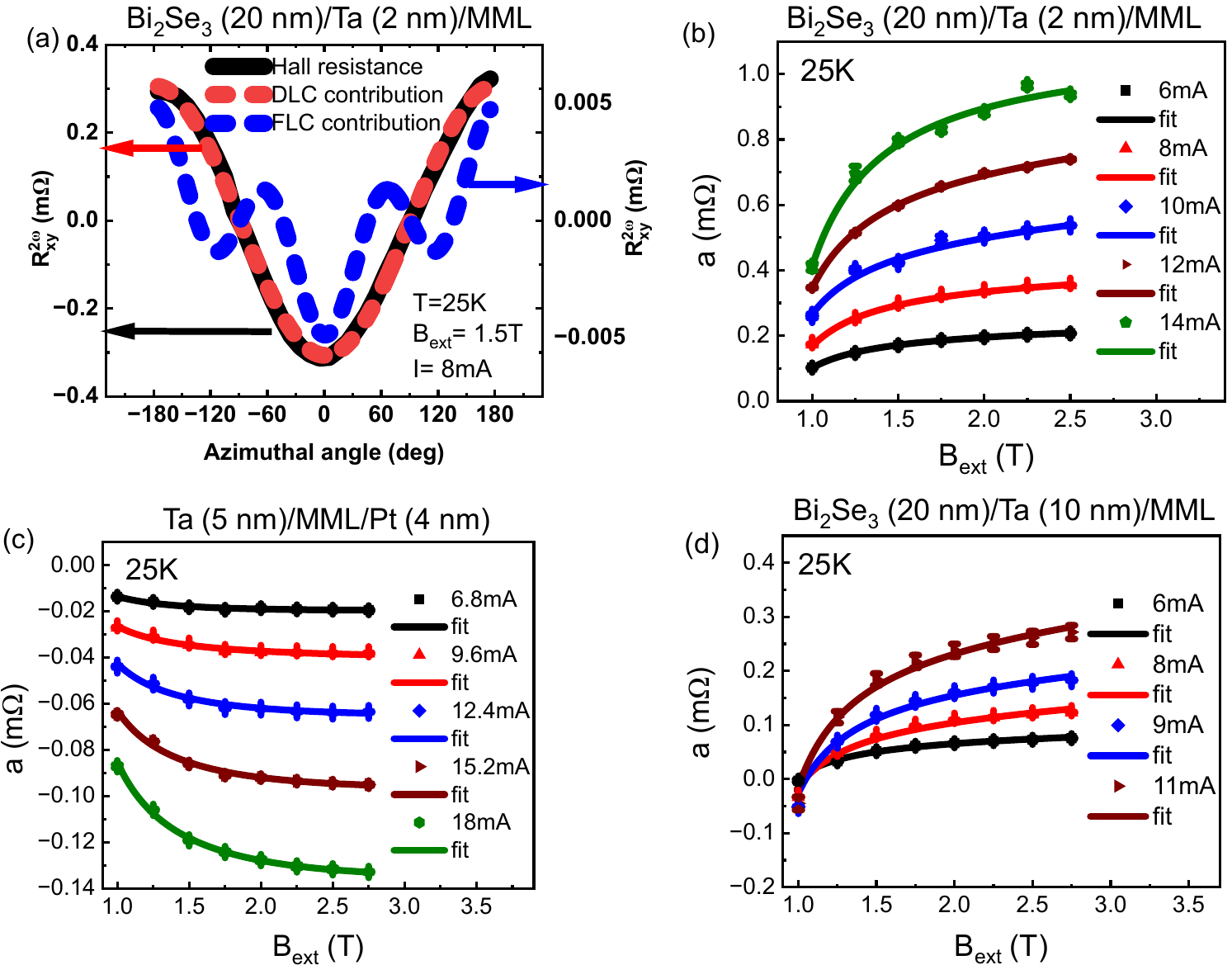}
	\caption{Second harmonic Hall measurements. (a) Representative azimuthal-angle-dependent second-harmonic Hall signals for the  Al$_2$O$_3$/Bi$_2$Se$_3$(20\,nm)/Ta(2\,nm)/MML measured at 25\,K, 1.5\,T, and an applied current of $\sim$9\,mA. Black points denote the experimental data, while the red and blue dashed lines represent contributions to a fit to Eq.~\ref{eq:angdep}, from which the damping-like and field-like contributions (DLC and FLC) are extracted. Note the change of scale for the FLC contribution. (b--d) Magnetic-field dependence of DLC for the three samples, fitted using Eq.~\ref{eq:dlc} to obtain the damping-like torque $B_{\mathrm{DL}}$ and thermal contributions at different applied currents. \label{fig:secondharm}} \end{figure}

The spin--orbit torques (SOTs) generated in the magnetic heterostructures were quantitatively examined using the second-harmonic Hall measurement technique, a widely established method \cite{manchon2019current,zhang2020topological,hayashi2014quantitative,pandey2025tunable,li2025oxidation,zhao2025field,roschewsky2019spin,ahn2023observation,du2021origin} for determining current-induced effective fields and separating the different SOT contributions, as explained in detail in the Supplementary Information. The SOT-induced effective field can be decomposed into two orthogonal components: a damping-like effective field $B_\mathrm{DL}$ (damping-like contribution, DLC) and a field-like effective field $B_\mathrm{FL}$ (field-like contribution, FLC). In addition, the second-harmonic signal may include thermoelectric contributions such as the ordinary Nernst effect (ONE), anomalous Nernst effect (ANE), spin Seebeck effect (SSE), and other thermal voltages arising from Joule heating and temperature gradients within the sample \cite{roschewsky2019spin,zhang2020topological,ahn2023observation,garello2013symmetry,yun2025quantitative}. The second-harmonic Hall resistance $R^{2 \omega}_{xy}$ can therefore be expressed as \cite{roschewsky2019spin,ahn2023observation,du2021origin}
\begin{equation}
R_{xy}^{2\omega} = -a \cos\phi +  b\left( 2\cos^{3}\phi - \cos\phi \right), \label{eq:angdep}
\end{equation}
where the coefficients $a$ and $b$ represent the DLC and FLC, respectively, written as
\begin{align}
a &= R_{\mathrm{AHE}} \frac{B_{\mathrm{DL}}}{B_{\mathrm{ext}} - B_{\mathrm{K}}}
+ C_{1} B_{\mathrm{ext}}
+ C_{2}, \label{eq:dlc} \\
b &= 2R_{\mathrm{PHE}} \frac{B_{\mathrm{FL+Oe}}}{B_{\mathrm{ext}}}. \label{eq:flc}
\end{align}
Here, $C_{1}$ represents the contribution proportional to the external field $B_{\mathrm{ext}}$ (e.g.\ from the ordinary Nernst effect, ONE),  and $C_{2}$ accounts for field-independent thermoelectric voltages arising from the anomalous Nernst effect (ANE) and/or the spin Seebeck effect (SSE). Also, $\phi$ is the azimuthal angle of the applied magnetic field $B_\mathrm{ext}$ (magnetization) with the applied current (see Fig.~\ref{fig:schematic}(b)), and $B_{\mathrm{FL+Oe}}$ is field-like (including Oersted) effective field, $B_{\mathrm{ext}}$ is the external magnetic field, and $B_{\mathrm{K}}$ is the anisotropy field, which we assume to be equal to the in-plane saturation field.

To separate the different contributions to the second-harmonic Hall response, an in-plane magnetic field $B_\mathrm{ext}$ with magnitude significantly larger than the anisotropy field $B_{\mathrm{K}}$ was applied while rotating its direction azimuthally over $360^{\circ}$, as illustrated in Fig.~\ref{fig:schematic}(b). Owing to their distinct angular symmetries, the damping-like-related and field-like-related contributions can be separated by fitting the measured $R_{xy}^{2\omega}(\phi)$ curves, yielding the coefficients associated with the damping-like (\(a\)) and field-like (\(b\)) symmetry channels, as shown in Fig.~\ref{fig:secondharm}(a). The field-like effective field, $B_{\mathrm{FL}}$, is directly obtained from the field dependence of the \(b\) coefficient. In contrast, the \(a\) coefficient contains contributions from both the damping-like torque and thermoelectric effects. These two contributions are subsequently separated based on their different dependences on the applied magnetic field \cite{roschewsky2019spin,ahn2023observation}, enabling the extraction of the damping-like effective field, $B_{\mathrm{DL}}$, and the thermoelectric component. While the damping-like and field-like torque contributions are discussed here, details of the thermoelectric contributions are provided in the Supplementary Information.

The current-induced Oersted field was estimated using the simplified expression \cite{chen2017quantifying,he2020study}
\begin{equation}
B_{\mathrm{Oe}} = \frac{\mu_{0} I}{2W}, \label{eq:oersted}
\end{equation}
where $I$ is the applied current and $W$ is the width of the Hall bar. Subtracting $B_{\mathrm{Oe}}$ from the extracted field-like term yielded the intrinsic field-like SOT. However, in our experimental samples, the $B_{\mathrm{Oe}}$ values are found to be in the same order as the observed field-like torque, which suggests that the observed field-like torque is mainly the Oersted field generated due to the charge current through the TI and HM layers. As can be seen from Fig.~\ref{fig:secondharm}(a), the FLC is much smaller than the DLC in our measurements, and we will not discuss it further. 

\begin{figure}
    \centering
	\includegraphics[width=5.5in]{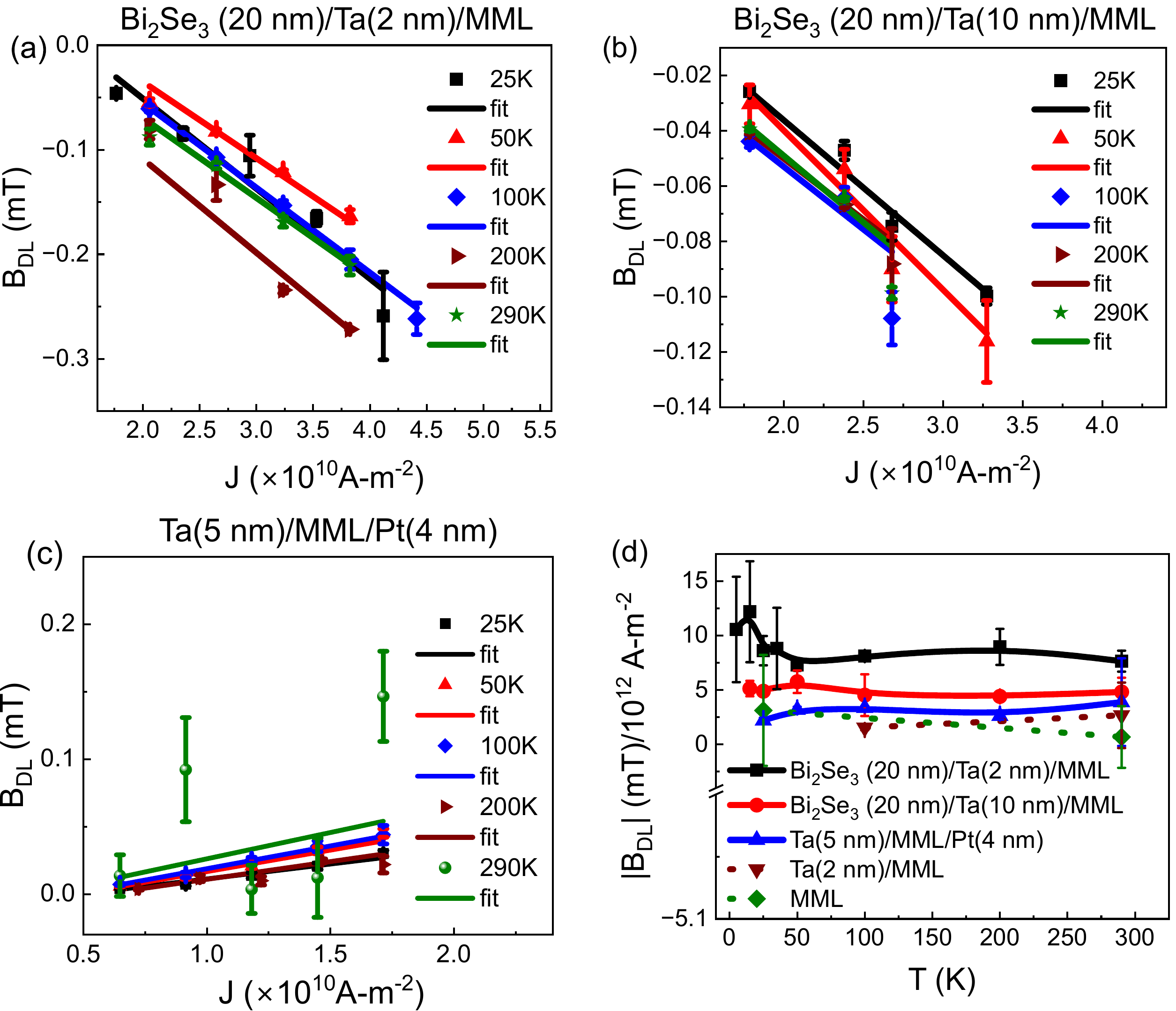}
    \caption{Damping-like contributions to the SOT. (a--c) Linear variation of the damping-like effective field $B_{\mathrm{DL}}$ with applied current density for the Al$_2$O$_3$/Bi$_2$Se$_3$(20\,nm)/Ta(2\,nm)/MML,  Al$_2$O$_3$/Bi$_2$Se$_3$(20\,nm)/Ta(10\,nm)/MML, and SiO$_2$/Ta(5\,nm)/MML/Pt(4\,nm) samples, respectively, measured at different temperatures. The points represent the $B_{\mathrm{DL}}$ values extracted from the DLC-versus-field fits, while the solid lines are linear fits used to determine the damping-like torque per unit current density. (d) Temperature dependence of the extracted $B_{\mathrm{DL}}$ per unit current density for all samples. Over the full temperature range, the damping-like torque per unit current density exhibits a clear hierarchy, with Al$_2$O$_3$/Bi$_2$Se$_3$(20\,nm)/Ta(2\,nm)/MML showing the largest values, followed by Al$_2$O$_3$/Bi$_2$Se$_3$(20\,nm)/Ta(10\,nm)/MML, and finally SiO$_2$/Ta(5\,nm)/MML/Pt(4\,nm). The two reference samples Ta(2\,nm)/MML and MML also exhibits  a little $B_{\mathrm{DL}}$ of nearly 3~mT/$10^{12}$~Am$^{-2}$. \label{fig:dlc}} 
\end{figure}

The value of $B_\mathrm{DL}$ at a given value of $J$ and $T$ was determined by fitting Eq.~\ref{eq:dlc} to the $B_\mathrm{ext}$ dependence of $a$. Examples for three different samples at 25~K for a range of values of $J$ are shown in Fig.~\ref{fig:secondharm}(b--d). The $J$ dependence of $B_\mathrm{DL}$ derived from these fits is shown in Fig.~\ref{fig:dlc}(a--c).

It is clearly visible in Fig.~\ref{fig:dlc}(a--c) that the damping-like fields exhibit an approximately linear dependence on the applied current density, consistent with their origin in spin--orbit torque (SOT) mechanisms. It is important to note, however, that although the effective fields scale linearly with current density, the corresponding linear fits do not necessarily pass through the origin. Instead, a finite intercept, corresponding to a threshold or critical current density required to generate a measurable SOT response, is frequently observed \cite{heinen2010current,fukami2013depinning}. Representative examples illustrating this behaviour are presented and discussed in supplementary information.

Performing these measurements over temperatures ranging from 25~K to 290~K provides a comprehensive view of the temperature dependence of the SOTs in the investigated heterostructures. The temperature dependence of $B_\mathrm{DL}$ for a selection of samples is shown in Fig.~\ref{fig:dlc}(d).

\section*{Discussion and Conclusion}

Second-harmonic Hall measurements across the studied heterostructures reveal several consistent trends that clarify the role of topological insulators in generating efficient spin--orbit torques (SOTs). The most striking observation is the substantial enhancement of the damping-like torque in samples incorporating Bi$_2$Se$_3$. Among all the samples, the samples with thinner Ta buffer, i.e. Al$_2$O$_3$/Bi$_2$Se$_3$(20,nm)/Ta(2,nm)/MML structure exhibits the largest torque magnitude, reaching approximately 8~mT/$10^{12}$ A,m$^{-2}$ down to 25 K, followed by a pronounced increase to approximately 12~mT/$10^{12}$ A,m$^{-2}$ at 15 K, together with a sharp change in behaviour at lower temperatures (Fig.~\ref{fig:dlc}(d)). In contrast, the SiO$_2$/Ta(5,nm)/MML/Pt(4,nm) sample, despite containing two heavy metals capable of generating spin currents, produces significantly smaller torque values of approximately 2.5~mT/$10^{12}$ A,m$^{-2}$. For the SiO$_2$/MML and SiO$_2$/Ta(2,nm)/MML samples, only a small damping-like torque of approximately 1--2~mT/$10^{12}$ A,m$^{-2}$ is observed. This is expected given the absence of an efficient SOT source in the SiO$_2$/MML structure and the small fraction of current (approximately 8--10\%) flowing through the thin Ta layer in the SiO$_2$/Ta(2,nm)/MML sample. Taking current shunting into account, the measured torque values are comparable to those reported for similar ferromagnet/non-magnet heterostructures in the literature \cite{han2023gate,zhang2020topological,zhang2024large,gamou2019enhancement,dai2021controllable,qiu2014angular}.

Another notable feature of the TI-based structures is the reversal of the damping-like torque polarity relative to samples without Bi$_2$Se$_3$. This reversed polarity persists even in the Bi$_2$Se$_3$(20,nm)/Ta(10,nm)/MML sample, although with a reduced magnitude of approximately 5~mT/$10^{12}$ A,m$^{-2}$. The persistence of the polarity reversal is noteworthy because the 10 nm Ta buffer layer would be expected to attenuate TI-generated spin currents before they reach the magnetic multilayer. Our results instead indicate that a substantial spin current originating from the TI still reaches the MML, in addition to the torque generated by Ta itself. This behaviour suggests a more complex mechanism governing spin-current transmission within the multilayer stack. One possible explanation is that the ultrathin Pt layer (approximately 1 nm) within the MML modifies or filters the torque generated by Ta, whose spin Hall angle has the opposite sign to that of Pt. Since Bi$_2$Se$_3$-derived torques are generally expected to have the same sign as Pt-generated torques, the observed polarity trends may reflect a competition between TI-, Ta-, and Pt-generated spin currents. A detailed microscopic investigation of this polarity behaviour is beyond the scope of the present work and will be the subject of future studies.

It is important to note that the torque efficiencies reported here are based on the \emph{total} applied current density. Only $\sim$30--35\% of this current flows through the Bi$_2$Se$_3$ layer when using the thinner 2\,nm Ta buffer, implying that the intrinsic SOT efficiency of the TI is significantly larger than the nominal values extracted from the device-level current density. Furthermore, while the Ta buffer layer attenuates a considerable portion of the TI-generated spin current, it remains essential for stabilizing the perpendicular magnetic anisotropy of the MML \cite{brereton2026tailoring}. This highlights the delicate balance between magnetic functionality and spin-transport transparency in complex heterostructures.

Despite this attenuation, the damping-like torque in the TI-based sample with the thin 2~nm Ta buffer remains remarkably large across the entire temperature range. Quantitatively, the DLT increases from approximately 8\,mT/ $10^{12}$\,A\,m$^{-2}$ at room temperature to nearly 12\,mT/ $10^{12}$\,A\,m$^{-2}$ at low temperatures. For the Bi$_2$Se$_3$/Ta(10\,nm)/MML sample, the corresponding values lie around $\sim$5\,mT/ $10^{12}$\,A\,m$^{-2}$, whereas the all-heavy-metal Ta/MML/Pt structure exhibits only $\sim$2.5\,mT/ $10^{12}$\,A\,m$^{-2}$. These comparisons clearly demonstrate the superiority of TI-driven torques and the substantial enhancement in torque strength achieved by incorporating Bi$_2$Se$_3$ into the heterostructure. When compared with the single Ta(2\,nm) reference sample, the enhancement of the damping-like torque due to the TI is unambiguous. Although similar trends are observed in the field-like torque, the FLC contains an additional contribution from the Oersted field generated by the charge current, and the measured values closely match the expected Oersted field magnitude.

For the TI samples with the thinner Ta buffer, a distinct temperature dependence is observed. At lower temperatures,  the DLT increases sharply. This behaviour likely reflects the combined influence of the emergence of topological surface states (TSS) and the increased current shunting through the Bi$_2$Se$_3$ layer as the temperature decreases. While one may argue that the enhancement can be explained solely by current shunting, it is important to note that the shunting increases continuously across the temperature range, whereas the DLT enhancement becomes pronounced only below $\sim$20\,K. In contrast, for all other samples, the DLT values remain nearly constant throughout the temperature range, further highlighting the unique role of the TI.

Overall, our results demonstrate that TI-derived spin--momentum--locked currents dominate the SOT response, particularly at low temperatures, and can be engineered through careful control of buffer-layer thickness, interfacial transparency, and stacking order. Beyond their fundamental significance, the exceptionally large torques observed here hold strong promise for technological applications, including energy-efficient magnetization switching, skyrmion or domain-wall manipulation, and next-generation memory or computing architectures. Future studies combining TI-based torques with engineered chiral textures, ultrathin magnetic layers, voltage-controlled interfaces, or non-linear spin-transport regimes may unlock even more efficient and multifunctional spintronic platforms.

\section*{Methods}

As described above, the investigated heterostructures consist of 20\, nm epitaxial Bi$_2$Se$_3$ layers, a prototypical topological insulator (TI), onto which a 
$[\mathrm{Pt}(0.8\,\mathrm{nm})/\mathrm{CoB}(0.6\,\mathrm{nm})/\mathrm{Ru}(0.5\,\mathrm{nm})] \times 6$ 
multilayer was grown on a Ta buffer layer. The TI films were grown by molecular beam epitaxy (MBE), whereas the multilayer stack was deposited by dc magnetron sputtering at room temperature \cite{brereton2026tailoring}.

The Bi$_2$Se$_3$ (20\,nm) layers were grown in an MBE chamber with a base pressure of approximately $7 \times 10^{-10}$\,mbar. Bismuth and selenium were co‑evaporated from Knudsen cells onto $c$‑plane sapphire substrates under Se‑rich conditions, with the selenium flux maintained at least twenty times higher than the bismuth flux to suppress Se vacancies. Growth proceeded in two steps: first, $\sim$2\,nm nucleation layer was deposited at $130^{\circ}\mathrm{C}$, followed by the remaining thickness grown at $300^{\circ}\mathrm{C}$, resulting in terraces approximately 100\,nm wide.

The MBE and sputtering chambers were interconnected, enabling the complete heterostructure to be fabricated without breaking ultrahigh vacuum. This ensured a clean TI surface and high‑quality interface. After transfer to the sputtering chamber, the Ta buffer layer and the multilayer stack were deposited at room temperature, and all samples were capped with 0.5~nm Ru.

\subsection*{Sample fabrication}

Hall-bar devices with various dimensions were fabricated using standard photolithography. A positive photoresist (S1813) was patterned using a maskless lithography system (Heidelberg Instruments: MLA150), followed by ion milling (Scia Mill 150) to define the device geometry. A second lithography step was then used to form the contact metallization through e-beam evaporation,depositing Ti (5\,nm)/Au (50\,nm) contacts.

\subsection*{Measurements}

All transport measurements were carried out in a cryogenic environment with a split-pair magnet to allow for sample rotation in the applied field (details provided in the Supplementary Information). A constant AC current was supplied using a Keithley 6221 current source, and the longitudinal and/or transverse voltage signals both first and second‑harmonic components, were detected using two independent lock‑in amplifiers. Magnetization measurements were performed using a SQUID‑VSM. Complementary MOKE measurements were conducted using an Evico wide‑field Kerr microscope system.

\section*{Data availability}
The data associated with this paper are openly available at the Research Data Leeds repository at DOI TBC.

\section*{Code availability}
Custom code used in this study is available from the corresponding author
upon reasonable request.

\section*{Acknowledgements}
S.H., A.Y., S.S., G.B. and C.H.M. acknowledge support from the EPSRC Programme Grant `CAMIE' (EP/X027074/1),  A.Y. and S.S. were also supported by the EPSRC Programme Grant 'NAME' (EP/V001914/1). The samples were grown in the Royce Deposition system at the University of Leeds, which is supported by the Henry Royce Institute, United Kingdom, through Grants No. EP/P022464/1 and No. EP/R00661X/1. 

\section*{Author contributions}
C.H.M. conceived and supervised the overall project. A.Y. carried out the growth of the $\text{Bi}_2\text{Se}_3$ topological insulator layers under the supervision of S.S., while S.H. and B.B. deposited the remaining multilayer stack under the supervision of C.H.M. S.H. and B.B. performed all device fabrication. Transport measurements were conducted by S.H., with assistance from G.B. for cryostat operation and software handling. All data analysis was performed by S.H., and the results were shared with and discussed among all authors.

\section*{Competing interests}
The authors declare no competing interests.

\section*{Additional information}
Supplementary information is available for this paper.

\bibliography{references}

\end{document}